\documentclass[aps, prstab, reprint, showpacs, amsmath, amssymb]{revtex4-1}

\usepackage{amsmath,amscd}
\usepackage{bm}% bold math
\usepackage[english]{babel}
\usepackage{graphicx}% Include figure files
\usepackage{soul}
\usepackage[usenames]{color}
\DeclareMathOperator{\w}{\omega}
\DeclareMathOperator{\ve}{\varepsilon}
\DeclareMathOperator{\R}{\rho}

\begin{document}
\title{ Wakefields in Hollow Channel of Magnetized Plasma }

\author{Sergey N. Galyamin}
\email{s.galyamin@spbu.ru}
%\author{Andrey V. Tyukhtin}
\affiliation{Saint Petersburg State University, 7/9 Universitetskaya nab., St. Petersburg, 199034 Russia}
%\author{Viktor V. Vorobev}
%\affiliation{Technical University of Munich, Germany, Department of Informatics}

\date{\today}

\begin{abstract}
Wakefield particle acceleration in hollow plasma channels is under extensive study nowadays. 
Here we consider an externally magnetized plasma layer (external magnetic field of arbitrary magnitude is along the structure axis) and investigate wakefields generated by a point charge passing along the layer axis.
\end{abstract}

\maketitle

\section{Introduction}
In recent years, an essential progress has been achieved in particle acceleration within Plasma Wakefield Acceleration (PWFA) scheme.
One of the most promising configuration of plasma -- hollow plasma channel -- has been investigated in a series of papers.
For example, in~\cite{Gessner16} both electron and positron acceleration with gradients that are orders of magnitude larger than those achieved in conventional accelerators has been shown.

%%%%%%%%%%%%%%%%%%%%%%%%%%%%%%%%%%%%%%%%%%%%%%%%%%%%%%%%%%%%%%%%%%%%%%%%%%%%%%%
\section{Problem formulation}
%%%%%%%%%%%%%%%%%%%%%%%%%%%%%%%%%%%%%%%%%%%%%%%%%%%%%%%%%%%%%%%%%%%%%%%%%%%%%%%

%
%%%%%%%%%%%%%%%%%%%%%%%%%%%%%%
\begin{figure}[b]
\centering
\includegraphics[width=0.4\textwidth]{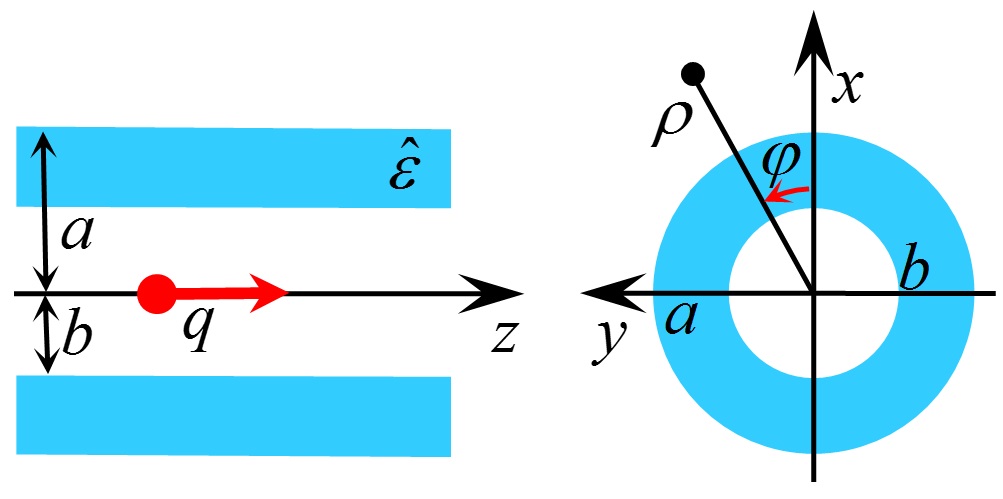}
\caption{\label{fig:geom} Geometry of the problem and main notations.}
\end{figure}
%%%%%%%%%%%%%%%%%%%%%%%%%%%%%%
%

We consider a layer of cold magnetized plasma with the inner radius $b$ and the outer radius $a$,~Fig.~\ref{fig:geom}.
Plasma is described by the following tensor of dielectric permittivity:
\begin{equation}
\hat{\ve}
=
\left(
\begin{array}{ccc}
{\ve_{\bot}}&{-ig}&0 \\
{ig}&{\ve_{\bot}}&0 \\
0&0&{\ve_{\parallel}}
\end{array}
\label{eq:tensor}
\right)
\end{equation}
and magnetic permeability
$\mu{=}1$.
The
$g$
component causes gyrotropy, while the inequality of components
$\ve_{\bot}$
and 
$\ve_{\parallel}$
creates uniaxial anisotropy.
Here,
$\ve_{\bot}$,
$g$
and
$\ve_{\parallel}$
are assumed to be frequency dependent, i.e., frequency dispersion is taken into account.
A model for a cold electron plasma in an external magnetic field is utilized here.
The frequency dependence of the components of the permittivity tensor in such a medium is described by the following expressions
\cite{Gb, GKT13}:
\begin{equation}
\begin{aligned}
&\ve_{\bot}(\w)
=
1 - \frac{\w_p^2 (\w + i\nu)}{\w \left[ (\w + i\nu)^2 - \w_h^2 \right]}, \\
&g(\w)
=
\frac{- \w_p^2{\w_h}}{\w \left[ (\w + i\nu)^2 - \w_h^2 \right]}, \\
&\ve_{\parallel}(\w)
= 1 - \frac{\w_p^2}{\w^2 + i\w\nu},
\end{aligned}
\label{eq:dispersion}
\end{equation}
where
$\w_p^2{=}4\pi Ne^2{/}m$
is the plasma frequency 
($N$
is the electron density, and
$e$
and
$m$
are the electron charge and the electron mass, respectively),
$\w_h{=}|e|H_{ext}{/}(mc)$
is a ``gyrofrequency'' 
($H_{ext}$
is the external magnetic field directed along $z$-axis) and
$\nu$
is the effective collision frequency.
The following analytical calculations are performed for the case where the medium has no losses 
($\nu{=}0$), 
but an infinitely small value of
$\nu$
can be used, if needed, to determine the branches of the radicals and the positional relationship of the integration path and the singularities.
Both the region outside the layer and the region inside the layer (channel) are supposed to be vacuum.

We assume that the charge moves with constant velocity $V$ along $z$-axis of the cylindrical frame $\R$,
$\varphi$,
$z$.
Thus, the charge and current densities have the form
\begin{equation}
\R^{\text{q}} 
= 
q\delta(\R)\delta(z - \upsilon t),
\qquad
\vec{j} = V \R^{\text{q}} \vec{e}_z,
\label{eq:source}
\end{equation}
$V = \beta c$, $c$ is the light speed in vacuum.
For the Gaussian bunch with the rms half-length
$ \sigma $,
\begin{equation}
\label{eq:gaussian}
\eta_{ \mathrm{ G } }( z - V t )
=
\frac{ 1 }{ \sqrt{ 2 \pi } \sigma }
\exp{ \left( \frac{ - ( z - V t )^2 }{ 2 \sigma^2 } \right) },
\end{equation}
it is sufficient to substitute
\begin{equation}
\label{eq:substg}
q
\to
q \exp{ \left( - \frac{ \omega^2 }{ \omega_{ \sigma }^2 } \right) },
\quad
\omega_{ \sigma } = \frac{ \sqrt{ 2 } V }{ \sigma }.
\end{equation}
in final expressions.

Note that the electromagnetic (EM) problem is further solved in the frequency domain so that Fourier integral decomposition is used. 
For example, longitudinal field components $E_z$ and $H_z$ have the form:
\begin{equation}
\left\{
E_z,
\,\,
H_z
\right\}
=
\int\nolimits_{-\infty}^{+\infty } \left\{ \tilde{ E }_{ \omega z }, \,\, \tilde{ H }_{\w z} \right\} e^{ -i \omega t }\,d\omega.
\end{equation}
Fourier magnitudes 
$ \tilde{ E }_{ \omega z } $ and $ \tilde{ H }_{\w z} $
are further used as ``potentials'' since the boundary value problem is solved for these components while the rest of components is expressed through them.  

%%%%%%%%%%%%%%%%%%%%%%%%%%%%%%%%%%%%%%%%%%%%%%%%%%%%%%%%%%%%%%%%%%%%%%%%%%%%%%%
\subsection{Field representation in vacuum areas}
%%%%%%%%%%%%%%%%%%%%%%%%%%%%%%%%%%%%%%%%%%%%%%%%%%%%%%%%%%%%%%%%%%%%%%%%%%%%%%%

First, let us consider the area (1), $0<\R<b$.
As it is known (see, for example,~\cite{GVT19}), self-field of a moving charge (this is an incident field for our boundary problem) can be written in the following form:
\begin{equation}
\label{eq:E0}
\begin{aligned}
\tilde{E}_{\w z}^{(i)}
&= E_0 K_0(\R \sigma_0) \exp{ \left( \frac{ i k_0 }{ \beta } z \right) } = E_{\w z}^{(i)} \exp{ \left( \frac{ i k_0 }{ \beta } z \right) }, \\
\tilde{H}_{\w z}^{(i)} &= 0,
\end{aligned}
\end{equation}
where
$ E_0 = q \sigma_0^2 / ( i \pi \w ) $,
$ \sigma_0^2 = k_0^2 / \beta^2 ( 1 - \beta^2 ) = k_0^2 / \beta^2 / \gamma^2 $
($\gamma$ is a Lorentz factor),
$ \sigma_0 = k_0 / \beta / \gamma $,
$K_0$ is a MacDonald function.

It is clear from~\eqref{eq:E0} that for the given geometry all field components should depend on $z$ in the same manner, i.e. $\sim \exp{ ( i k_0 z / \beta ) } $.
In other words, it is natural to suppose that Fourier magnitudes for field components in all areas of the problem should have the following form:
\begin{equation}
\label{eq:expiomegav}
\left\{ \tilde{ E }_{ \omega z }(\R,z), \,\, \tilde{ H }_{\w z}(\R,z) \right\}
=
\left\{ E_{ \omega z }(\R), \,\, H_{\w z}(\R) \right\}
\exp( i k_0 z / \beta),
\end{equation}
Therefore, the scattered field for $ 0 < \R < b $ is expressed as follows:
\begin{equation}
\label{eq:EHchannel}
E_{\w z}^{(1)}
= A I_0(\R \sigma_0),
\quad
H_{\w z}^{(1)} = B I_0(\R \sigma_0),
\end{equation}
where $A$ and $B$ are unknown constants,
$I_0$ is a modified Bessel function.
Note that while deriving~\eqref{eq:EHchannel} we have utilized the following facts: $ E_{\w z} $ and $ H_{\w z} $ in vacuum satisfy the modified Bessel equation of the 0-th order and scattered field should be free of singularity for $\R \to 0$~\cite{GVT19}. 

Similar considerations lead to the following representation in the region outside the plasma layer, area (3), $ \R > a $:
\begin{equation}
\label{eq:EHoutside}
E_{\w z}^{(3)}
= D K_0(\R \sigma_0),
\quad
H_{\w z}^{(3)} = F K_0(\R \sigma_0),
\end{equation}
where $D$ and $F$ are unknown constants.
Other field components are expressed as follows:
\begin{equation}
\label{eq:componentsvac}
\begin{aligned}
H_{\w \varphi}^{(1,3)} &= \frac{ i }{ k_0 } \frac{ \partial E_{\omega z}^{(1,3)} }{ \partial \R }, & 
E_{\w \R}^{(1,3)} &= \frac{ H_{\w \varphi}^{(1,3)} }{ \beta }, \\
E_{\w \varphi}^{(1,3)} &= \frac{ 1 }{ i k_0 } \frac{ \partial H_{\omega z}^{(1,3)} }{ \partial \R }, &
H_{\w \R}^{(1,3)} &= - \frac{ E_{\w \varphi}^{(1,3)} }{ \beta }.
\end{aligned}
\end{equation}
In the issue, EM field in vacuum areas is determined by four unknowns, $A$, $B$, $D$ and $F$.

%%%%%%%%%%%%%%%%%%%%%%%%%%%%%%%%%%%%%%%%%%%%%%%%%%%%%%%%%%%%%%%%%%%%%%%%%%%%%%%
\subsection{Field representation in plasma layer}
%%%%%%%%%%%%%%%%%%%%%%%%%%%%%%%%%%%%%%%%%%%%%%%%%%%%%%%%%%%%%%%%%%%%%%%%%%%%%%%

Maxwell equations without external sources in plasma medium~\eqref{eq:tensor} 
\begin{equation}
\label{Maxwell}
\begin{aligned}
\mathrm{rot}\vec{\tilde{E}}_{\w} &= i k_0 \vec{\tilde{E}}_{\w}, &
\quad
\mathrm{rot}\vec{\tilde{H}}_{\w} &= - i k_0 \vec{\tilde{D}}_{\w}, \\
\mathrm{div}\vec{\tilde{H}}_{\w} &= \mathrm{div}\vec{\tilde{B}}_{\w} = 0, &
\quad
\mathrm{div}\vec{\tilde{D}}_{\w} &= 0,
\end{aligned}
\end{equation}
with material relations 
\begin{equation}
\label{eq:matrel}
\begin{aligned}
\vec{D}_{\w}&=\hat{\ve}\vec{E}_{\w}= \vec{e}_{z} \ve_{\parallel} E_{\w z} +\\ 
&+ \vec{e}_{\R}[\ve_{\bot} E_{\w \R} - i g E_{\w \varphi} ] + 
\vec{e}_{\varphi}[\ve_{\bot} E_{\w \varphi} + i g E_{\w \R} ],
\end{aligned}
\end{equation}
and~\eqref{eq:expiomegav} applied result in the following system of two coupled equations for $E_{\omega z}^{(2)}$ and $H_{\omega z}^{(2)}$ 
\begin{equation}
\label{eq:coupled}
\begin{aligned}
\left(
\frac{ \partial^2 }{ \partial \R^2 } + \frac{ 1 }{ \R } \frac{ \partial }{ \partial \R } + k_E^2 
\right)
E_{\omega z}^{(2)} &= -i \alpha^2 H_{\omega z}^{(2)}, \\
\left(
\frac{ \partial^2 }{ \partial \R^2 } + \frac{ 1 }{ \R } \frac{ \partial }{ \partial \R } + k_H^2 
\right)
H_{\omega z}^{(2)} &= i \alpha^2 \ve_{ \parallel } E_{\omega z}^{(2)},
\end{aligned}
\end{equation}
where
\begin{equation}
\label{eq:design}
\begin{aligned}
k_E^2 &= k_0^2 \ve_{ \parallel } - \frac{ k_0^2 \ve_{ \parallel } }{ \beta^2 \ve_{ \bot } },
\quad
k_H^2 = k_0^2 \frac{ \ve_{ \bot }^2 - g^2  }{ \ve_{ \bot } } - \frac{ k_0^2 }{ \beta^2 }, \\
\alpha^2 &= \frac{ k_0^2 g }{ \ve_{ \bot } \beta }.
\end{aligned}
\end{equation}
Other components are expressed as follows:
\begin{equation}
\label{eq:componentsplasma}
\begin{aligned}
H_{\w \varphi} &= \frac{ \left[ \ve_{ \bot } \left( \ve_{ \bot } - \beta^{ -2 } \right) - g^2 \right]
\dfrac{ \partial E_{\omega z}^{(2)} }{ \partial \R } -
\dfrac{ i g }{ \beta } \dfrac{ \partial H_{\omega z}^{(2)} }{ \partial \R } }
{ i k_0 \left[  g^2 - \left( \ve_{ \bot } - \beta^{ -2 } \right)^2 \right] }, \\
E_{\w \R} &= - \frac{ \dfrac{ i }{ \beta } \left( \ve_{ \bot } - \beta^{ -2 } \right) \dfrac{ \partial E_{\omega z}^{(2)} }{ \partial \R } +
g \dfrac{ \partial H_{\omega z}^{(2)} }{ \partial \R } }
{ k_0 \left[  g^2 - \left( \ve_{ \bot } - \beta^{ -2 } \right)^2 \right] },\\
E_{\w \varphi} &= \frac{ i\left( \ve_{ \bot } - \beta^{ -2 } \right) \dfrac{ \partial H_{\omega z}^{(2)} }{ \partial \R } -
\dfrac{ g }{ \beta } \dfrac{ \partial E_{\omega z}^{(2)} }{ \partial \R } }
{ k_0 \left[  g^2 - \left( \ve_{ \bot } - \beta^{ -2 } \right)^2 \right] }, \\
H_{\w \R} &= - \frac{ E_{\w \varphi} }{ \beta }.
\end{aligned}
\end{equation}
Note that~\eqref{eq:componentsvac} is a particular case of \eqref{eq:componentsplasma} for $g=0$, $\ve_{ \bot } = \ve_{ \parallel } = 1$.

The system~\eqref{eq:coupled} can be solved, for example, as follows. 
Let us use the substitution
\begin{equation}
\label{eq:uncouple}
E_{\omega z}^{(2)}( \R ) = C_E H_0^{(1)}(\xi),
\quad
H_{\omega z}^{(2)}( \R ) = C_H H_0^{(1)}(\xi),
\end{equation}
resulting in homogeneous system for $C_E$ and $C_H$.
Corresponding determinant should be equal to zero, therefore
\begin{equation}
\label{eq:disp}
\left( \xi^2 - k_E^2 \right)
\left( \xi^2 - k_H^2 \right)
- \ve_{ \parallel } \alpha^4 = 0.
\end{equation}
Solutions of Eq.~\eqref{eq:disp} $\xi^2 = s_{o,e}^2$ are squared transverse wave numbers of the ordinary ($s_o$) and the extraordinary ($s_e$) waves~\cite{Gb, GKT13}:
\begin{equation}
\label{eq:soe}
\begin{aligned}
s_{o,e}^{2}
&{=}
\frac{ k_0^2 }{ 2 \beta^2 \ve_{\bot} }
\left[
\vphantom{
\sqrt{ \left[ ( \ve_{\bot}^2 - g^2 - \ve_{ \bot } \ve_{ \parallel } ) \beta^2 - \ve_{ \bot } + \ve_{ \parallel } \right]^2
+
4 \beta^2 g^2 \ve_{ \parallel } }
}
( \ve_{\bot}^2 - g^2 +\ve_{ \bot } \ve_{ \parallel } ) \beta^2 - \ve_{ \bot } - \ve_{ \parallel } \right.
\pm \\
&\left.
\sqrt{ \left[ ( \ve_{\bot}^2 - g^2 - \ve_{ \bot } \ve_{ \parallel } ) \beta^2 - \ve_{ \bot } + \ve_{ \parallel } \right]^2
+
4 \beta^2 g^2 \ve_{ \parallel } }
\right],
\end{aligned}
\end{equation}
where ``--'' corresponds to ``o'' and ``+'' corresponds to ``e''.
Since we have 4 solutions for $\xi$, $\xi = \pm s_o$, $\xi = \pm s_e$, then the general solution for \eqref{eq:coupled} can be expressed as follows:
\begin{equation}
\label{eq:gensolE1}
\begin{aligned}
E_{\omega z}^{(2)} & = C_{E+}^{o} H_0^{(1)}(\R s_o) + C_{E-}^{o} H_0^{(2)}(\R s_o) + \\
& + C_{E+}^{e} H_0^{(1)}(\R s_e) + C_{E-}^{e} H_0^{(2)}(\R s_e),
\end{aligned}
\end{equation}
where $C_{E\pm}^{o,e}$ some constants.
This solution can be equivalently rewritten as follows:
\begin{equation}
\label{eq:gensolE2}
\begin{aligned}
E_{\omega z}^{(2)} &= C_1^{o} J_0(\R s_o) + C_2^{o} N_0(\R s_o) + \\
& + C_1^{e} J_0(\R s_e) + C_2^{e} N_0(\R s_e),
\end{aligned}
\end{equation}
where
$ C_1^{o} $, $ C_2^{o} $, $ C_1^{e} $ and $ C_2^{e} $ are other unknown constants that should be determined. 

For $H_{\omega z}$ we have (corresponding constants can be expressed by substitution of \eqref{eq:gensolE2} to \eqref{eq:coupled}): 
\begin{equation}
\label{eq:gensolH}
\begin{aligned}
H_{\omega z}^{(2)} & = C_1^{o} i \frac{ k_E^2 - s_o^2 }{ \alpha^2 } J_0(\R s_o) + 
C_2^{o} i \frac{ k_E^2 - s_o^2 }{ \alpha^2 } N_0(\R s_o) + \\
& + C_1^{e} i \frac{ k_E^2 - s_e^2 }{ \alpha^2 } J_0(\R s_e) + 
C_2^{e} i \frac{ k_E^2 - s_e^2 }{ \alpha^2 } N_0(\R s_e).
\end{aligned}
\end{equation}
In the issue, EM field in plasma is determined by four unknowns, $ C_1^{o} $, $ C_2^{o} $, $ C_1^{e} $ and $ C_2^{e} $.

%%%%%%%%%%%%%%%%%%%%%%%%%%%%%%%%%%%%%%%%%%%%%%%%%%%%%%%%%%%%%%%%%%%%%%%%%%%%%%%
\section{General solution}
%%%%%%%%%%%%%%%%%%%%%%%%%%%%%%%%%%%%%%%%%%%%%%%%%%%%%%%%%%%%%%%%%%%%%%%%%%%%%%%

We introduce a vector $\vec{X}$ of unknowns:
\begin{equation}
\label{eq:X}
\vec{X} = \left(
\begin{array}{cccccccc}
A & B & C_1^{o} & C_1^{e} & C_2^{o} & C_2^{e} & D & F 
\end{array}
\right)^{\mathrm{T}}.
\end{equation}
Continuity of tangential EM field components
\begin{equation}
\label{eq:continuity}
\begin{aligned}
E_{\omega z}^{(i)}(b) + E_{\omega z}^{(1)}(b) &= E_{\omega z}^{(2)}(b), & 
H_{\omega z}^{(1)}(b) &= H_{\omega z}^{(2)}(b), \\
E_{\omega \varphi}^{(i)}(b) + E_{\omega \varphi}^{(1)}(b) &= E_{\omega \varphi}^{(2)}(b), & 
H_{\omega \varphi}^{(1)}(b) &= H_{\omega \varphi}^{(2)}(b), \\
E_{\omega z}^{(2)}(a) &= E_{\omega z}^{(3)}(a), & 
H_{\omega z}^{(2)}(b) &= H_{\omega z}^{(3)}(b), \\
E_{\omega \varphi}^{(2)}(a) &= E_{\omega \varphi}^{(3)}(a), & 
H_{\omega \varphi}^{(2)}(a) &= H_{\omega \varphi}^{(3)}(a),
\end{aligned}
\end{equation}
results in the following linear system for $\vec{X}$:
\begin{equation}
\label{eq:MXF}
\hat{M}\vec{X} = \vec{F}
\end{equation}
where
\begin{widetext}
\begin{equation}
\label{eq:M}
\hat{M} =
\left(
\begin{array}{cccccccc}
-I_0( b \sigma_0 ) & 0 & J_0(bs_o) & J_0(bs_e) & N_0(bs_o) & N_0(bs_e) & 0 & 0 \\
0 & -I_0(b\sigma_0) & i\frac{k_E^2-s_o^2}{\alpha^2}J_0(bs_o) & i\frac{k_E^2-s_e^2}{\alpha^2}J_0(bs_e) & i\frac{k_E^2-s_o^2}{\alpha^2}N_0(bs_o) & i\frac{k_E^2-s_e^2}{\alpha^2}N_0(bs_e) & 0 & 0 \\
\frac{ik_0}{\sigma_0}I_1(b\sigma_0) & 0 & \frac{is_o h_o}{k_0 f \beta^2}J_1(bs_o) & \frac{is_e h_e}{k_0 f \beta^2}J_1(bs_e) &
\frac{is_o h_o}{k_0 f \beta^2}N_1(bs_o) & \frac{is_e h_e}{k_0 f \beta^2}N_1(bs_e) & 0 & 0 \\
0 & \frac{k_0}{i\sigma_0}I_1(b\sigma_0) & \frac{s_o d_o}{k_0f\beta^2}J_1(bs_o) & \frac{s_e d_e}{k_0f\beta^2}J_1(bs_e) &
\frac{s_o d_o}{k_0f\beta^2}N_1(bs_o) & \frac{s_e d_e}{k_0f\beta^2}N_1(bs_e) & 0 & 0 \\
0 & 0 & J_0(as_o) & J_0(as_e) & N_0(as_o) & N_0(as_e) & -K_0(a\sigma_0) & 0 \\
0 & 0 & i \frac{k_E^2 - s_o^2}{\alpha^2}J_0(as_o) & i \frac{k_E^2 - s_e^2}{\alpha^2}J_0(as_e) & i \frac{k_E^2 - s_o^2}{\alpha^2}N_0(as_o) &
i \frac{k_E^2 - s_e^2}{\alpha^2}N_0(as_e) & 0 & -K_0(a\sigma_0) \\
0 & 0 & \frac{ i s_o h_o}{k_0 f \beta^2} J_1(as_o) & \frac{ i s_e h_e}{k_0 f \beta^2} J_1(as_e) &
\frac{ i s_o h_o}{k_0 f \beta^2} N_1(as_o) & \frac{ i s_e h_e}{k_0 f \beta^2} N_1(as_e) & \frac{k_0}{i\sigma_0}K_1(a\sigma_0) & 0 \\
0 & 0 & \frac{s_o d_o}{k_0 f \beta^2}J_1(as_o) & \frac{s_e d_e}{k_0 f \beta^2}J_1(as_e) &
\frac{s_o d_o}{k_0 f \beta^2}N_1(as_o) & \frac{s_e d_e}{k_0 f \beta^2}J_1(as_e) & 0 & \frac{ik_0}{\sigma_0}K_1(a\sigma_0)
\end{array}
\right),
\end{equation}
\end{widetext}
\begin{equation}
\label{eq:Y}
\vec{Y}=
\left(
\begin{array}{cccccccc}
-E_0 K_0( b \sigma_0 ) & 0 & E_0 \frac{ k_0 }{ i \sigma_0 } K_1( b \sigma_0 ) & 0 & 0 & 0 & 0 & 0
\end{array}
\right)^{\mathrm{T}},
\end{equation}
\begin{equation}
\begin{aligned}
f &=
k_0^2 \left[ g^2 - \left( \ve_{ \bot } - \beta^{ -2 } \right)^2 \right], \\
h_{o,e} &= \beta^2 \left[ \frac{ \ve_{\bot} }{ \beta^2 } \left( \ve_{ \bot } \beta^2 - 1 \right) -g^2 + 
\frac{g}{\beta}\frac{k_E^2 - s_{o,e}^2}{\alpha^2} \right], \\
d_{o,e} &= \frac{ k_E^2 - s_{o,e}^2 }{ \alpha^2 } \left( \ve_{ \bot } \beta^2 - 1 \right) + g \beta.
\end{aligned}
\end{equation}

Determinant $ \Delta $ of the matrix $ \hat{ M } $ can be calculated via two minors, $R_{11}$ and $R_{31}$:
\begin{equation}
\label{eq:det}
\Delta
=
-I_0( b \sigma_0 )R_{11} + \frac{ik_0}{\sigma_0}I_1(b\sigma_0)R_{31}.
\end{equation}
According to Kramer method~\cite{FMb}, 
\begin{equation}
\label{eq:AB}
A = \frac{\Delta_A}{\Delta},
\quad
B = \frac{\Delta_B}{\Delta},
\end{equation}
where the determinant $\Delta_A$ is expressed through the same minors because $\vec{Y}$ and the first column of $\hat{M}$ have the same structure: 
\begin{equation}
\label{eq:detA}
\Delta_A
=
-E_0 K_0( b \sigma_0 )R_{11} + \frac{E_0 k_0}{i\sigma_0}K_1(b\sigma_0)R_{31}.
\end{equation}
Calculation of $\Delta_B$ is a bit more complicated, for example:
\begin{equation}
\label{eq:detB}
\Delta_B
=
-I_0( b \sigma_0 )\tilde{R}_{11} + \frac{i k_0}{\sigma_0}I_1(b\sigma_0)\tilde{R}_{31},
\end{equation}
where columns $2-7$ of $\tilde{R}_{11}$ and $\tilde{R}_{31}$ coincide with those of $R_{11}$ and $R_{31}$, while the first column of $\tilde{R}_{11}$ and $\tilde{R}_{31}$ is, correspondingly:
\begin{equation}
\label{eq:1stcolumns}
\left(
\begin{array}{c}
0 \\
E_0 \frac{k_0}{i\sigma_0} K_1( b \sigma_0 ) \\
0 \\
0 \\
0 \\
0 \\
0
\end{array}
\right),
\quad
\left(
\begin{array}{c}
-E_0 K_0( b \sigma_0 ) \\
0 \\
0 \\
0 \\
0 \\
0 \\
0
\end{array}
\right).
\end{equation}
For practical purposes (to accelerate numerical calculations) it would be probably useful to express $R_{11}$ and $R_{31}$ through 48 $2 \times 2$ determinants (and do the same for $\tilde{R}_{11}$ and $\tilde{R}_{31}$), but corresponding bulky expressions are not shown here.

%%%%%%%%%%%%%%%%%%%%%%%%%%%%%%%%%%%%%%%%%%%%%%%
\section{Numerical results\label{sec:num}}
%%%%%%%%%%%%%%%%%%%%%%%%%%%%%%%%%%%%%%%%%%%%%%%

%
%%%%%%%%%%%%%%%%%%%%%%%%%%%%%%
\begin{figure}[]
\centering
\includegraphics[width=0.85\linewidth]{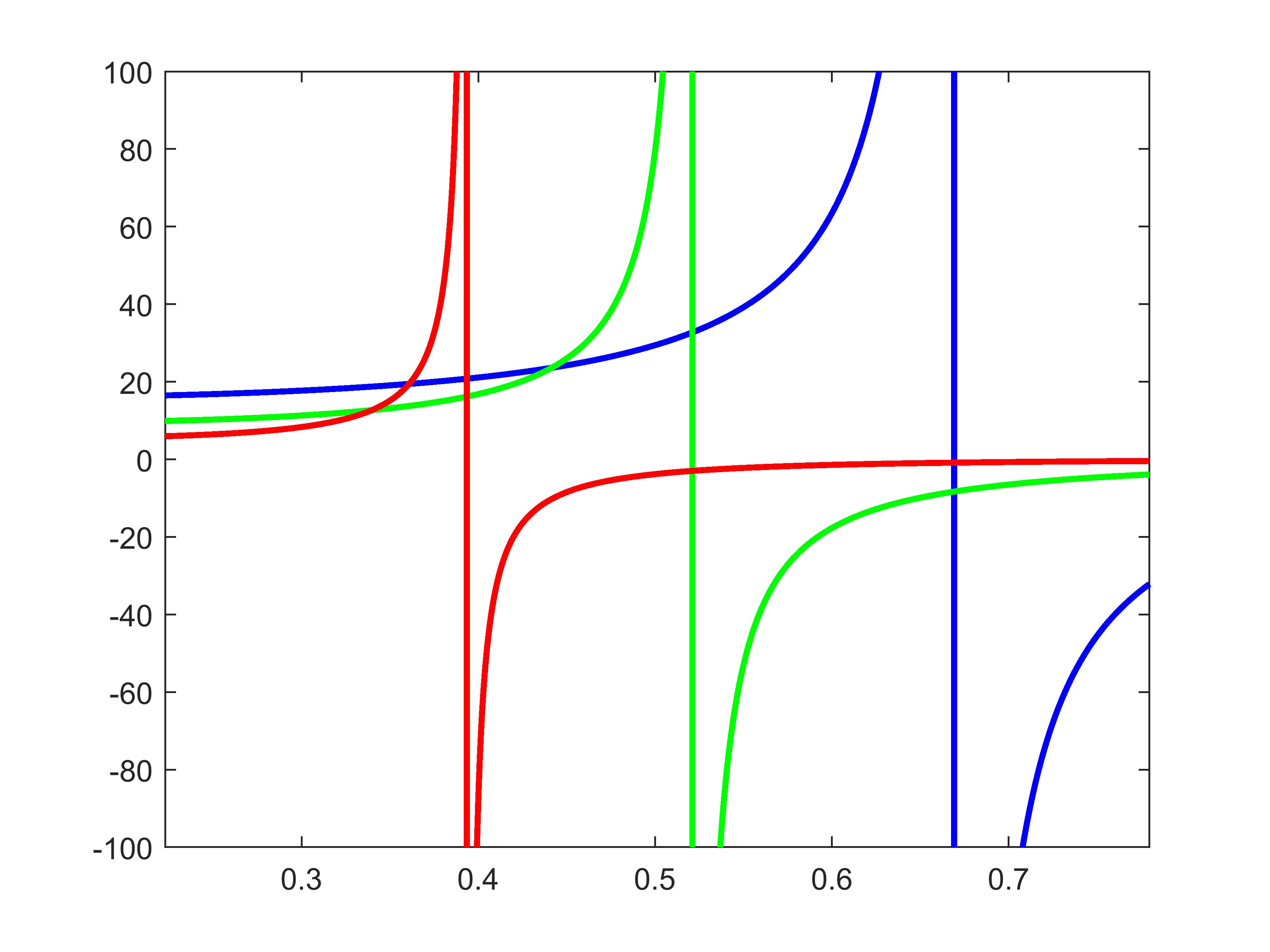}
	\caption{\label{fig:pole}Dependence of $\Delta_A / \Delta $ (arbitrary units) on $\w/\w_p$ for $a=10\lambda_p$ and $b=2\lambda_p$ (red), $b=\lambda_p$ (green) and $b=0.5\lambda_p$ (blue). Other parameters are: $q = -1$nC, $\beta = 0.9897$ ($\gamma = 7$), $\w_p = 10^{12}\mathrm{s}^{-1}$, $\w_h = 0.5\w_p$, $\nu = 0.001\w_p$, $\lambda_p = 2 \pi c /\w_p \approx 0.19$cm.}
\end{figure}
%%%%%%%%%%%%%%%%%%%%%%%%%%%%%%
%

Numerical calculations of the EM field distribution can be performed straightforwardly.
Since we are mainly interested in accelerating ($E_z$) and deflecting ($E_{\R}$) EM fields inside the channel, it is sufficient to obtain coefficient $A$ using Eq.~\eqref{eq:AB} and then calculate the inverse Fourier integrals for $E_z$ and $E_{\R}$. Corresponding calculations have been performed in Matlab using parallel computing toolbox.

According to~\cite{GKT13}, in the case of infinite plasma~\eqref{eq:tensor}, \eqref{eq:dispersion} a strong harmonic EM field is generated in the small neighborhood of the charge trajectory behind the charge. The components of most interest -- longitudinal ($E_z$) and transverse ($E_{\R}$) electric fields -- show singularities for $\R \to 0 $: $E_z$ a weak (logarithmic) singularity, while $E_{\R}$ a stronger (inversely proportional) singularity.
In general, these properties are similar to those of the ``plasma trace'', which is usually occurs in isotropic plasma
\begin{equation}
\label{isotropic} 
\ve(\w) = 1 - \w_p^2 / \w^2.
\end{equation}

An important difference between the plasma~\eqref{eq:tensor} and the isotropic plasma~\eqref{isotropic} consists in the fact that the magnitude of the orthogonal electric field decreases with an increasing external magnetic field $H_{ext}$ while magnitude of the longitudinal electric field does not depend on $H_{ext}$ and is determined by $\w_p$. 
In other words, $H_{ext}$ serves as additional parameter allowing suppressing the deflecting field.

When there is a vacuum channel in plasma~\eqref{isotropic}, the EM field of the described nature is absent inside the channel since there is no plasma here. However, it is known~\cite{Agadullin14} that for arbitrary frequency dispersive medium (and for plasma in particular) a specific surface wave is generated on the channel wall having the structure similar to that of ``plasma trace''. When channel radius tends to zero this surface wave transforms exactly to the ``plasma trace''. It is also clear that for enough small channel a charged particle passing along the channel axis will be strongly affected by these surface waves.     

Taking into account two aforementioned considerations, we can conclude that we should seek for a surface wave on the inner wall in the case under consideration. 

Figure~\ref{fig:pole} shows dependence $\Delta_A / \Delta $ (this ratio defines $E_z$ and $E_{\R}$ inside the channel, see Eqs.~\eqref{eq:AB} and \eqref{eq:EHchannel}) on $\w$. One can clearly see that this function possesses a singularity for some $\w$ depending on problem parameters. This is probably a pole singularity which we are seeking for.
Further we will perform EM field calculations for gaussian bunch~\eqref{eq:gaussian}.
Therefore in order to observe the contribution of this pole in the EM field, the length of the bunch $\sigma$ should be chosen small enough so that $\omega_{ \sigma } $ will be of order or less then  the pole singularity frequency.   

%
%%%%%%%%%%%%%%%%%%%%%%%%%%%%%%
\begin{figure*}[]
\centering
\includegraphics[width=0.7\linewidth]{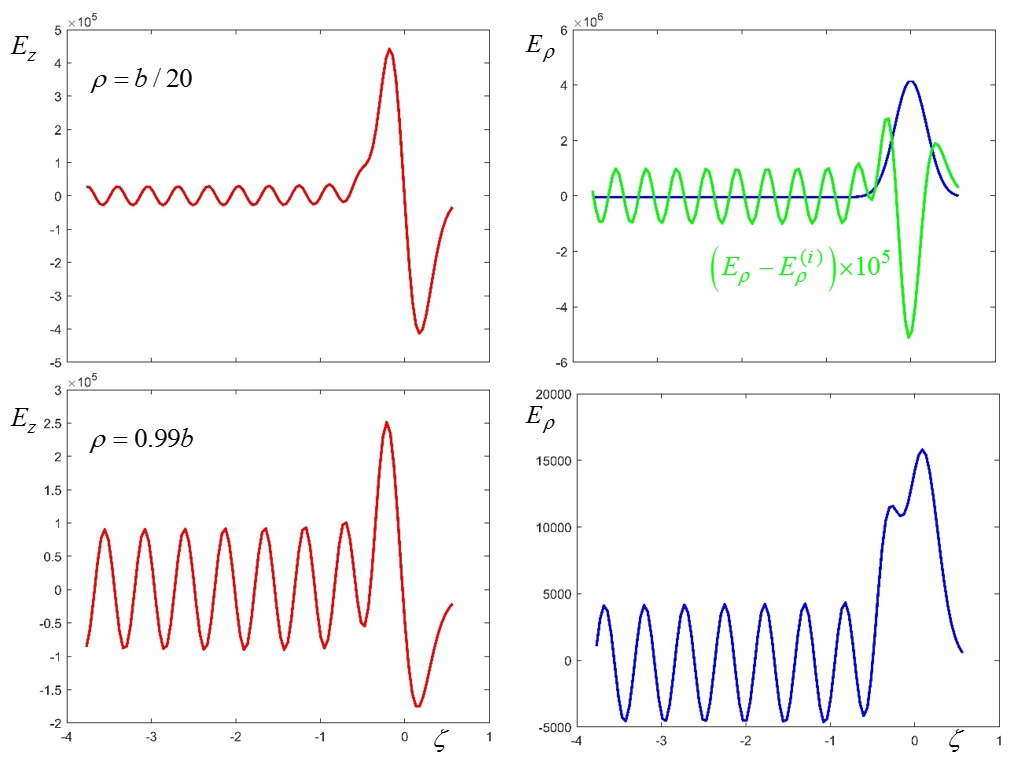}
	\caption{\label{fig:fields}Dependence of $E_z$ and $E_{\R}$~($\mathrm{V}/\mathrm{m}$) on $\zeta = z - \beta c t $ ($ \zeta \in [-20\lambda_p, 3\lambda_p]$, $\zeta$ in cm) for $a=10\lambda_p$ and $b=2\lambda_p$ and $r$ close to the axis (top) and close to the channel wall (bottom). Gaussian bunch length $\sigma = \lambda_p$, $\beta = 0.9897$ ($\gamma = 7$). Other parameters are the same as in Fig.~\ref{fig:pole}.}
\end{figure*}
%%%%%%%%%%%%%%%%%%%%%%%%%%%%%%
%

%
%%%%%%%%%%%%%%%%%%%%%%%%%%%%%%
\begin{figure*}[]
\centering
\includegraphics[width=0.8\linewidth]{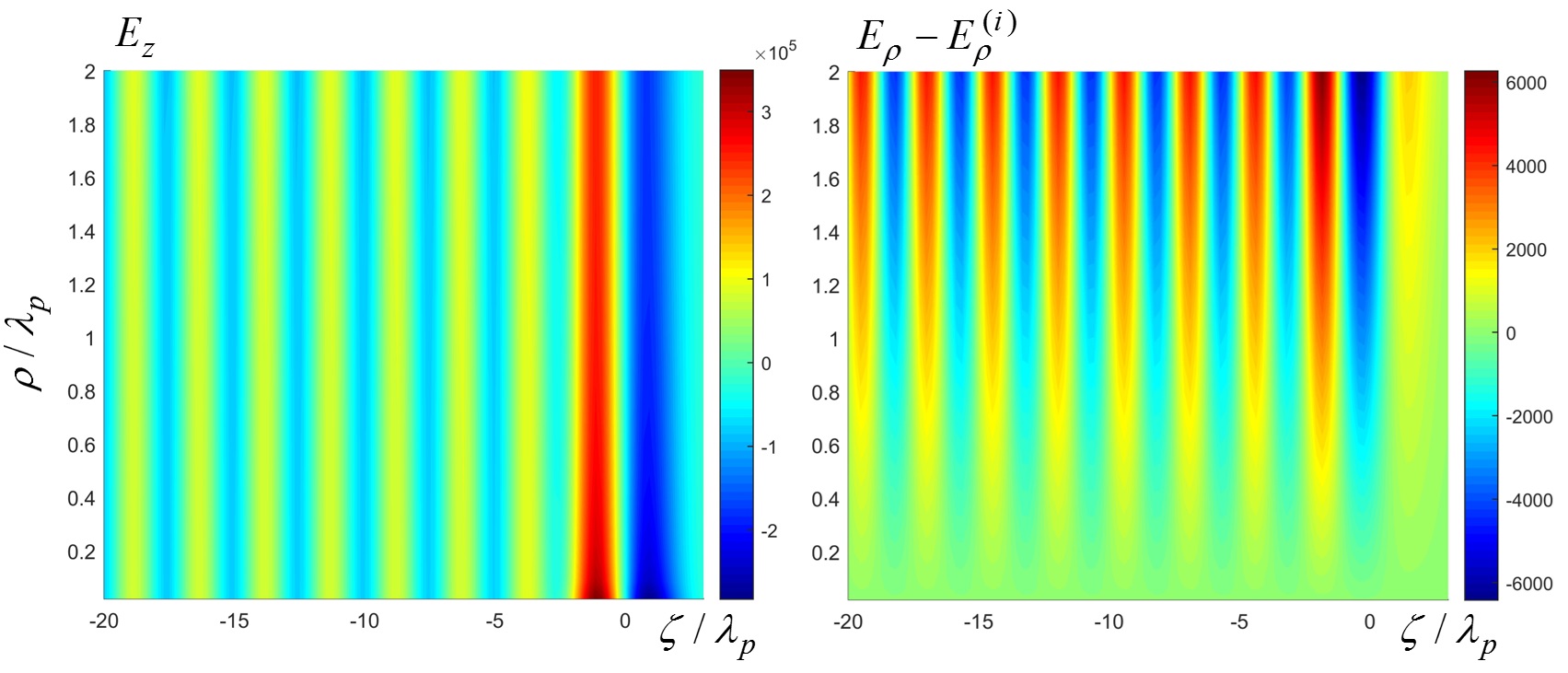}
	\caption{\label{fig:fields2D}Two-dimensional distribution of $E_z$ (total field) and $E_{\R} - E_{\R}^{(i)}$ (wave field only) (fields in $\mathrm{V}/\mathrm{m}$) for $a=10\lambda_p$ and $b=2\lambda_p$. Other parameters are the same as in Fig.~\ref{fig:fields}.}
\end{figure*}
%%%%%%%%%%%%%%%%%%%%%%%%%%%%%%
%

Figure~\ref{fig:fields} shows typical plots for longitudinal ($E_z$) and transverse ($E_{\R}$) field components near the channel axis and near the the channel wall.
First, one can see an expressed harmonic behavior behind the strong peak (which corresponds to the self-field of the gaussian bunch). The harmonic (single frequency) behavior indicates that this field originates from the pole contribution thus again proving our hypothesis discussed above. 
One can see that transverse field equals zero on the axis which is natural. 
Transverse field increases with $\R$ and possesses maximum for $\R=b$.
However, transverse field is always weaker than longitudinal, the latter depends weakly on $\R$.
This fact is again illustrated in Fig.~\ref{fig:fields2D} where two-dimensional distribution are shown. One can clearly see that there is always certain region near the channel axis where the transverse field can be neglected compared to the longitudinal one.

%
%%%%%%%%%%%%%%%%%%%%%%%%%%%%%%
\begin{figure*}[]
\centering
\includegraphics[width=0.95\linewidth]{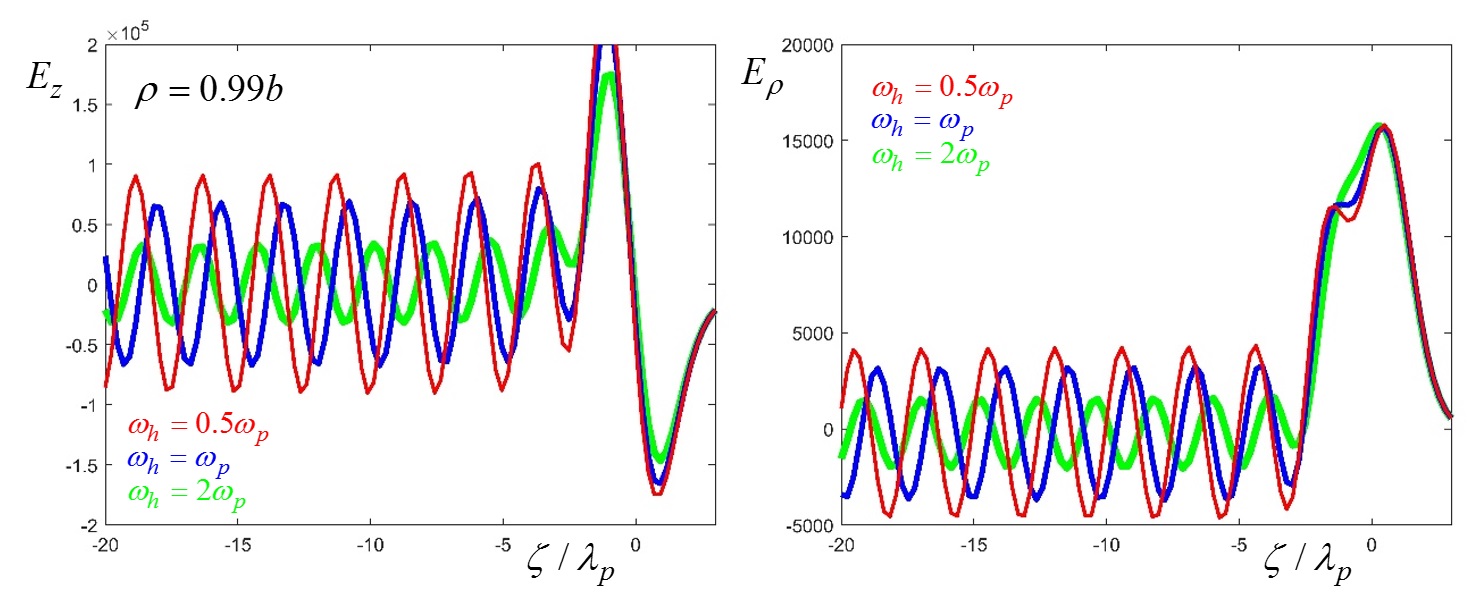}
	\caption{\label{fig:fieldswh} Dependence of $E_z$ and $E_{\R}$~($\mathrm{V}/\mathrm{m}$) on $\zeta = z - \beta c t $ ($\zeta$ in units of $\lambda_p$) for $a=10\lambda_p$, $b=2\lambda_p$ and $r$ close to the channel wall for various $\w_h$ indicated on plots. Gaussian bunch length $\sigma = \lambda_p$, $\beta = 0.9897$ ($\gamma = 7$). Other parameters are the same as in Fig.~\ref{fig:pole}.}
\end{figure*}
%%%%%%%%%%%%%%%%%%%%%%%%%%%%%%
%

It is also interesting to clarify how the external magnetic field $H_{ext}$ affects the discussed field distributions.
Figure~\ref{fig:fieldswh} shows typical field behavior near the channel wall for different $\w_h \sim H_{ext}$. One can see that magnitude of both longitudinal and transverse wave field decreases with an increase in $H_{ext}$ 

\section{Conclusion}

We have presented investigation of wakefields generated by a charged particle bunch (gaussian bunch has been used as a convenient example) inside a channel in a layer of cold magnetized electron plasma.
Such a layer of plasma is usually referred to as ``hollow plasma channel'' and extensively used in experiments on plasma wakefield acceleration (PWFA).
The used plasma model allows accounting for the external magnetization resulting in both anysotropy and gyrotropy of the medium.

We have constructed a rigorous solution of this boundary problem and have utilized it for development of effective numerical approach for calculation of transverse and longitudinal wakefields inside the channel.  
We have shown that longitudinal wakefields with magnitudes of order of $\mathrm{MV/m}$ (for plasma frequencies $\w_p \sim 10^{12}\mathrm{s}^{-1}$ and bunch charge $q \sim 1\mathrm{nC}$) can be generated and hollow plasma channel.
It is important that longitudinal EM field is practically uniform along the cross-section of the channel and typically much larger compared to the transverse one.
This is usually true even on the channel wall, where the transverse field is maximum, and even more so in the vicinity of the axis (approximately up to the half of channel radius).
Moreover, the external magnetization can be utilized for manipulation of the EM field structure in the channel.

\section{Acknowledgments}

This work was supported by Russian Science Foundation (Grant No.~18-72-10137).

%\bibliography{SNG_Bibliography_Apr2021}
%
\end{document}